\documentclass[useAMS,usenatbib]{mn2e}
\usepackage{natbib}
\usepackage{graphicx}\usepackage{epsfig}\usepackage{epsf}
\usepackage{amsmath}\usepackage{amssymb}\usepackage{stfloats}
\newcommand{\kms}{\ensuremath{{\rm km\,s^{-1}}}}                   
\newcommand{\msun}{\ensuremath{\mathit{M}_{\odot}}}               
\newcommand{\K}{\mathrm{K}}
\newcommand{\lsun}{\ensuremath{\mathit{L}_{\odot}}}                  
\newcommand{\rsun}{\ensuremath{\mathit{R}_{\odot}}}                  
\newcommand\ion[2]{#1$\;${\scshape{#2}}}%

\newcommand{\geff}{\ensuremath{\mathrm{g}_{\rm eff}}}                
\newcommand{\lstar}{\ensuremath{\mathit{L}_{\star}}}                 
\newcommand{\mdot}{\ensuremath{\dot{M}}}                             
\newcommand{\teff}{\ensuremath{\mathit{T}_{\rm eff}}}                
\newcommand{\vinf}{\ensuremath{v_{\infty}}}                          
\newcommand{\tauross}{\ensuremath{\tau_{\mathrm{Ross}}}}      

\title[SBW1's weak wind]{UV spectroscopy of the blue supergiant SBW1:
  the remarkably weak wind of a SN~1987A analog}

\author[Smith et al.]{Nathan Smith$^{1}$\thanks{E-mail:
    nathans@as.arizona.edu}, Jose H.\ Groh$^{2}$, Kevin France$^3$,
  and Richard McCray$^{4}$ \\ $^{1}$Steward Observatory, University of
  Arizona, 933 N. Cherry Ave., Tucson, AZ 85721, USA \\ $^2$Trinity
  College Dublin, The University if Dublin, Dublin 2, Ireland \\
  $^3$Laboratory for Atmospheric and Space Physics, University of
  Colorado, 600 UCB, Boulder, CO 80309, USA \\
  $^4$Dept.\ of Astronomy, University of California, Berkeley, CA
  94720-3411, USA}

\begin{document}

\pagerange{\pageref{firstpage}--\pageref{lastpage}} \pubyear{2015}
\maketitle
\label{firstpage}

\begin{abstract}

  The Galactic blue supergiant SBW1 with its circumstellar ring nebula
  represents the best known analog of the progenitor of
  SN~1987A. High-resolution imaging has shown H$\alpha$ and IR
  structures arising in an ionized flow that partly fills the ring's
  interior.  To constrain the influence of the stellar wind on this
  structure, we obtained an ultraviolet (UV) spectrum of the central
  star of SBW1 with the {\it Hubble Space Telescope} ({\it HST})
  Cosmic Origins Spectrograph (COS).  The UV spectrum shows none of
  the typical wind signatures, indicating a very low mass-loss rate.
  Radiative transfer models suggest an extremely low rate below
  10$^{-10}$ $M_{\odot}$ yr$^{-1}$, although we find that cooling
  timescales probably become comparable to or longer than the flow
  time below 10$^{-8}$ $M_{\odot}$ yr$^{-1}$. We therefore adopt this
  latter value as a conservative upper limit.  For the central star,
  the model yields $T_{\rm eff}$=21,000$\pm$1000~K,
  $L\simeq$5$\times$10$^4$ $L_{\odot}$, and roughly Solar composition
  except for enhanced N abundance. SBW1's very low mass-loss rate may
  hinder the wind's ability to shape the surrounding nebula.  The very
  low mass-loss rate also impairs the wind's ability to shed angular
  momentum; the spin-down timescale for magnetic breaking is more than
  500 times longer than the age of the ring.  This, combined with the
  star's slow rotation rate, constrain merger scenarios to form ring
  nebulae.  The mass-loss rate is at least 10 times lower than
  expected from mass-loss recipes, without any account of clumping.
  The physical explanation for why SBW1's wind is so weak presents an
  interesting mystery.
\end{abstract}


\section{INTRODUCTION}

SN~1987A was the nearest supernova (SN) in modern times. Two
surprising observations associated with SN~1987A (see review by
\citealt{arnett+89}) were the identification of a blue supergiant
(BSG) progenitor in pre-explosion images
\citep{walborn89,rousseau78,arnett87,arnett+89} and its very unusual
triple-ring circumstellar nebula \citep{burrows95,crotts95}.  These
two are intimately related, since the geometry of the nebula bears the
imprint of mass loss shaped by binary interaction and/or rapid
rotation as the star evolved to its blue pre-SN state.  The dynamical
age of the nebula is only $\sim$20,000 yr \citep{meaburn95,ch00}, so
the nebular structures trace recent pre-SN mass loss on a time scale
much shorter than core He burning and longer than C burning.  The
total mass of the ring is uncertain (due to an uncertain neutral
fraction), but may be 0.1-1 $M_{\odot}$ \citep{fransson15}.  An
important unresolved question is whether or not close binary evolution
was key in determining the progenitor's BSG state (mass transfer or
merger, mass loss, etc.).  The ring nebula may therefore provide
important clues to how and why the progenitor came to be a BSG.

Understanding the origin of the observed triple-ring structure has
been difficult, however.  Several early models showed that a faster
BSG wind expanding into a previous slower red supergiant (RSG) wind
with an equatorial density enhancement could yield an equatorial ring
and bipolar structure \citep{lm91,bl93,ma95,collins99}.  However, the
pair of polar rings around SN~1987A really are empty rings, rather
than limb-brightened polar lobes or filled caps, and their origin has
not been satisfactorally explained.  It is difficult to understand the
origin of the equatorial density enhancement in a RSG wind without
invoking a binary \citep{collins99}.  In subsequent studies, two
different types of models have been proposed as plausible ways to form
the nebula.  A scenario involving a binary merger as a RSG and
subsequent blueward evolution was proposed \citep{mp07,mp09}, and was
suggested to account for the observed nebular structure.  However,
this specific merger model predicts filled polar caps and relatively
empty mid-latitudes in the nebula, whereas the observed nebula has no
polar caps and may have some emission in the side walls of an
hourglass structure.  Moreover, the model requires that the merger
product should be rotating very rapidly, which seems to be at odds
with Galactic analogs (see below).  A somewhat different model
involves rotating single-star evolution, where a massive star spins up
as it contracts on a post-RSG blue loop, nearly reaching critical
rotation and ejecting a ring \citep{chita08}.  Then a bipolar wind
from the rotating BSG expands into the RSG wind and ring, forming
transient structures that may resemble the rings of SN~1987A
\citep{chita08}. However, this model also requires a rapidly rotating
BSG, inconsistent with observations of Galactic analogs. In either
case the strength of the BSG wind is a critical ingredient.

Radio observations of SN~1987A (or rather, the lack of radio emission
at early times) suggested that for the first 1000-1500 days after
explosion, the blast wave was expanding relatively unimpeded through a
very low-density wind \citep{ss93}.  After about 1500 days, however,
the radio emission brightened and the expansion speed slowed to only
3500 km s$^{-1}$ \citep{gaensler97,gaensler00,zanardo13}.  In order to
reach the angular scale of the resolved radio emission when it turned
on, the blast wave must have been expanding at about 35,000 km
s$^{-1}$ for those first 1500 days \citep{ss93}.  \citet{cd95}
suggested that after the initial free expansion through a rarefied
wind with $\dot{M}$=7.5$\times$10$^{-8} M_{\odot}$ yr$^{-1}$ or less,
the blast wave slammed into a dense H~{\sc ii} region that partly
filled the ring's interior.  In this model, the dense H~{\sc ii}
region was created by photoionization of the RSG wind by the BSG's UV
radiation.  This collision slowed the forward shock's expansion and
caused the radio emission to brighten dramatically \citep{cd95}.


Since the progenitor star Sk$-$69$^{\circ}$202 is now dead, it is hard
to improve our understanding of the connection between the star and
its nebula.  For this reason, nearby analogs of SN~1987A's progenitor
-- where the BSG has not yet exploded -- become interesting and
valuable.  There are currently three well-studied analogs known in our
Galaxy: (1) Sher~25 \citep{brandner97,smartt02}, which is
significantly more luminous and has partial polar caps instead of
rings, (2) HD~168625 \citep{smith07}, which is a more luminous LBV
candidate that does have polar rings, and (3) SBW1 \citep{sbw}
(discussed below).\footnote{Another possible member of the group is
  MN18, which is a similar BSG with a ring-like bipolar nebula
  \citep{gvaramadze15}, although this object has not yet been studied
  in as much detail as the others.}  An interesting recent result
places important constraints on the formation of these rings:
\citet{taylor14} monitored the central stars of all three Galactic
analogs with high-resolution spectroscopy and did not find any radial
velocity variations consistent with close massive binaries (the value
of sin $i$ is presumed from the resolved equatorial rings; if a binary
system's orbit were significantly misaligned with these rings, then
binary interation would not help to explain them).  Perhaps even more
interesing, \citet{taylor14} found that all three BSG central stars
have relatively {\it slow rotation speeds}.  For SBW1, the rotation
speed is only about 40 km s$^{-1}$.  Such slow rotation may be quite
problematic for some merger models that would predict a rapidly
rotating BSG post-merger product -- especially if the stellar wind in
the BSG phase is very weak.  We will return to this issue later in the
paper.

Of these three Galactic analogs, SBW1 is currently the best known
analog to SN~1987A in terms of stellar properties and nebular
structure.  The ring around SBW1 was discovered by \citet{sbw} during
a survey of the Carina Nebula, but two factors suggest that it is
actually located several kpc {\it behind} the Carina Nebula and is
seen there in projection.  First, its positive radial velocity
compared to expectations for Galactic rotation in that direction
suggest that it is on the far side of the Sagittarius-Carina arm and
outside the Solar circle.  Second, its apparent magnitude and color
only match its spectral type and luminosity class (B1.5 Iab;
\citealt{sbw}) if it is at a much larger distance than the Carina
Nebula.  At a distance of 6-7 kpc, the stellar luminosity
(0.5--1$\times$10$^5$ $L_{\odot}$) as well as the size of the ring
nebula ($r \simeq 0.2$ pc) make SBW1 a close match for SN~1987A.

Detailed analysis of the SBW1 nebula has provided interesting clues
that may alter our ideas about the formation of the nebula around
SN~1987A.  {\it HST} images of SBW1 \citep{smith13} show a pattern of
clumps around the ring and a radial extent that closely resemble the
spacing and size scale of spots in the ring of SN~1987A.  The interior
of the ring is filled with diffuse H$\alpha$ emission, although the
ring would probably appear much brighter relative to the interior if
it were flash ionized by a SN \citep{smith13}.  A very interesting
result is that high-resolution ground-based infrared (IR) images show
that the interior of the ring is also partly filled with diffuse
emission from warm dust.  Since BSGs don't produce dust in their
winds, this requires that the dust inside the ring was entrained from
the ring itself.  \citet{smith13} proposed that this structure arises
because the inner surface of the dense and neutral equatorial ring is
ionized by the central star, and that this triggers a dusty ionized
photoevaporative flow that fills the interior of the ring.  The
ionized gas expands into the ring until it collides with the stellar
wind; entrained dust piles up at this interface, producing the
observed peaks of thermal-IR emission inside the ring \citep{smith13}.
This simple ionized flow is able to dramatically influence the
observed structure and dynamics of the nebula because the ring's
expansion is slow (10-20 km s$^{-1}$), comparable to the sound speed
of the ionized gas.  This directly imaged structure around SBW1
appears to validate the picture of an H~{\sc ii} region inside the
ring of SN~1987A proposed by \citet{cd95}, which was deduced from the
time evolution of the SN's radio emission.  In this scenario, the main
requirement is that the star ejected a thin, dense ring about 10$^4$
yr ago.  While this might have occurred in post-RSG evolution to the
blue (perhaps with a merger), a ring might also be ejected in a brief
mass-transfer episode of a close binary or in an eruptive mass loss
event from a rotating star (e.g., \citealt{st07}).  A previous RSG
phase is not necessarily required \citep{sbw}.

\begin{figure*}
\includegraphics[width=\textwidth]{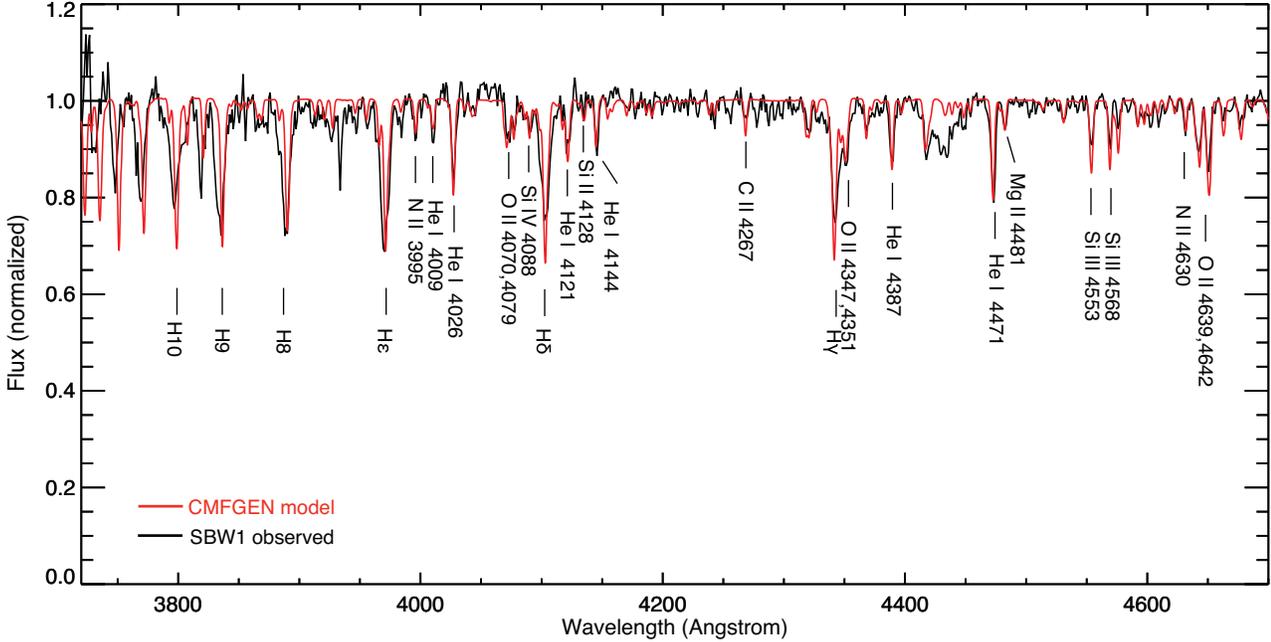}
\caption{\label{modeloptical} Comparison between the observed optical
  spectrum of SBW1 in the range $3760-4690$~\AA~(black line) and the
  best fit CMFGEN model (red line). The strongest spectral features
  are identified.  The broad feature at $\lambda\simeq$ 4430 \AA \ in
  the observed spectrum is due to absorption by a known diffuse
  interstellar band. There is a slight error in the wavelength
  solution at the blue end of the spectrum (note that Balmer lines are
  shifted slightly blueward compared to the model), due to poor signal
  to noise in the arc spectrum used for calibration. When we inferred
  \geff from this spectrum, we compared to individual lines with an
  appropriate shift for each.}
\end{figure*}

\begin{figure*}
\includegraphics[width=6in]{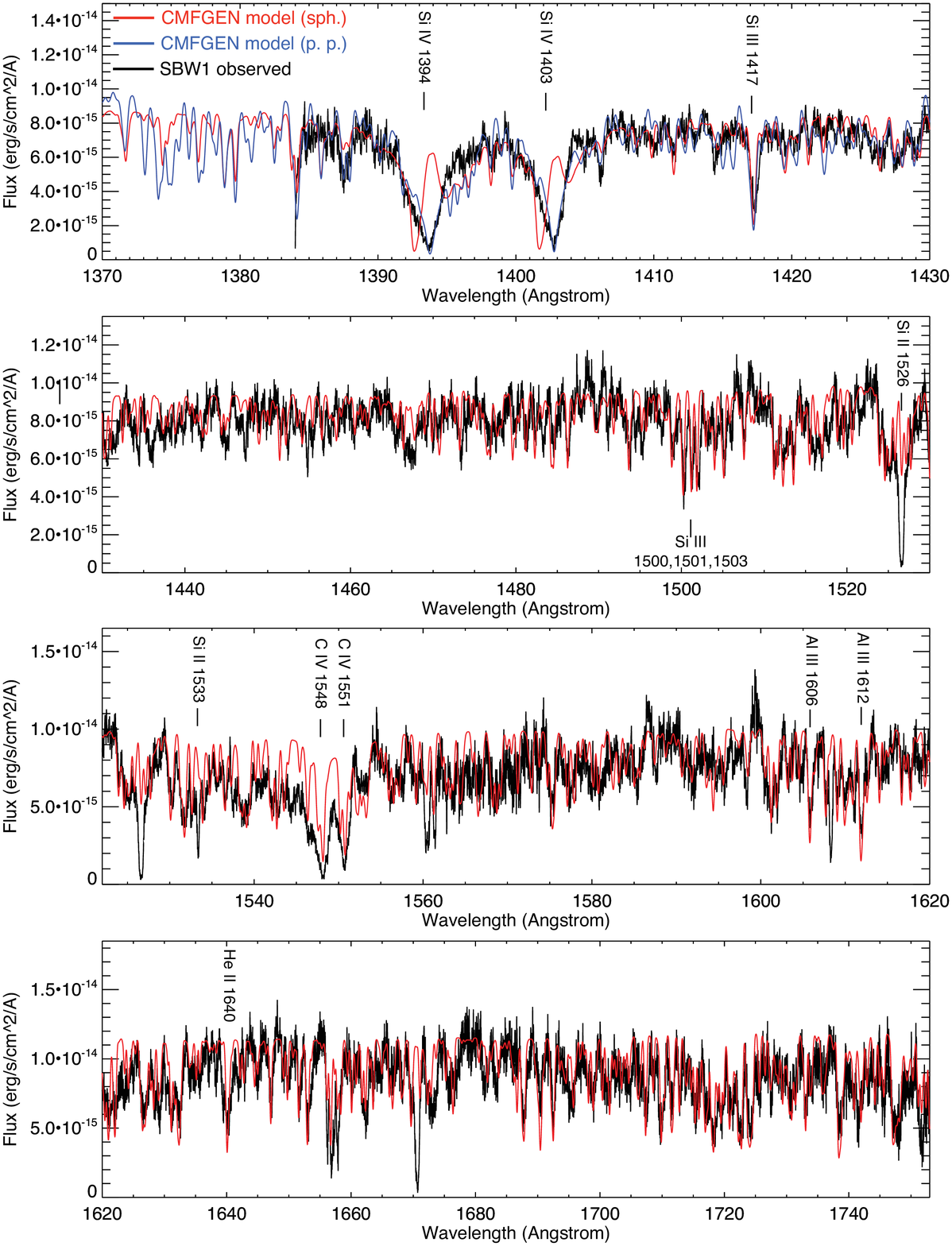}
\caption{\label{modeluv} Similar to Fig.~\ref{modeloptical}, but in
  the ultraviolet region between $1370-1750$~\AA. The strongest
  spectral lines are identified, while the remaining features are
  mostly due to \ion{Fe}{iii}, \ion{Fe}{iv}, or \ion{Ni}{iii} lines.
  This figure also includes a plane-parallel (p.p.) CMFGEN model in
  the UV (top panel, in blue) to illustrate the atmospheric spectrum
  with no wind.}
\end{figure*}

\begin{figure*}
\includegraphics[width=5.8in]{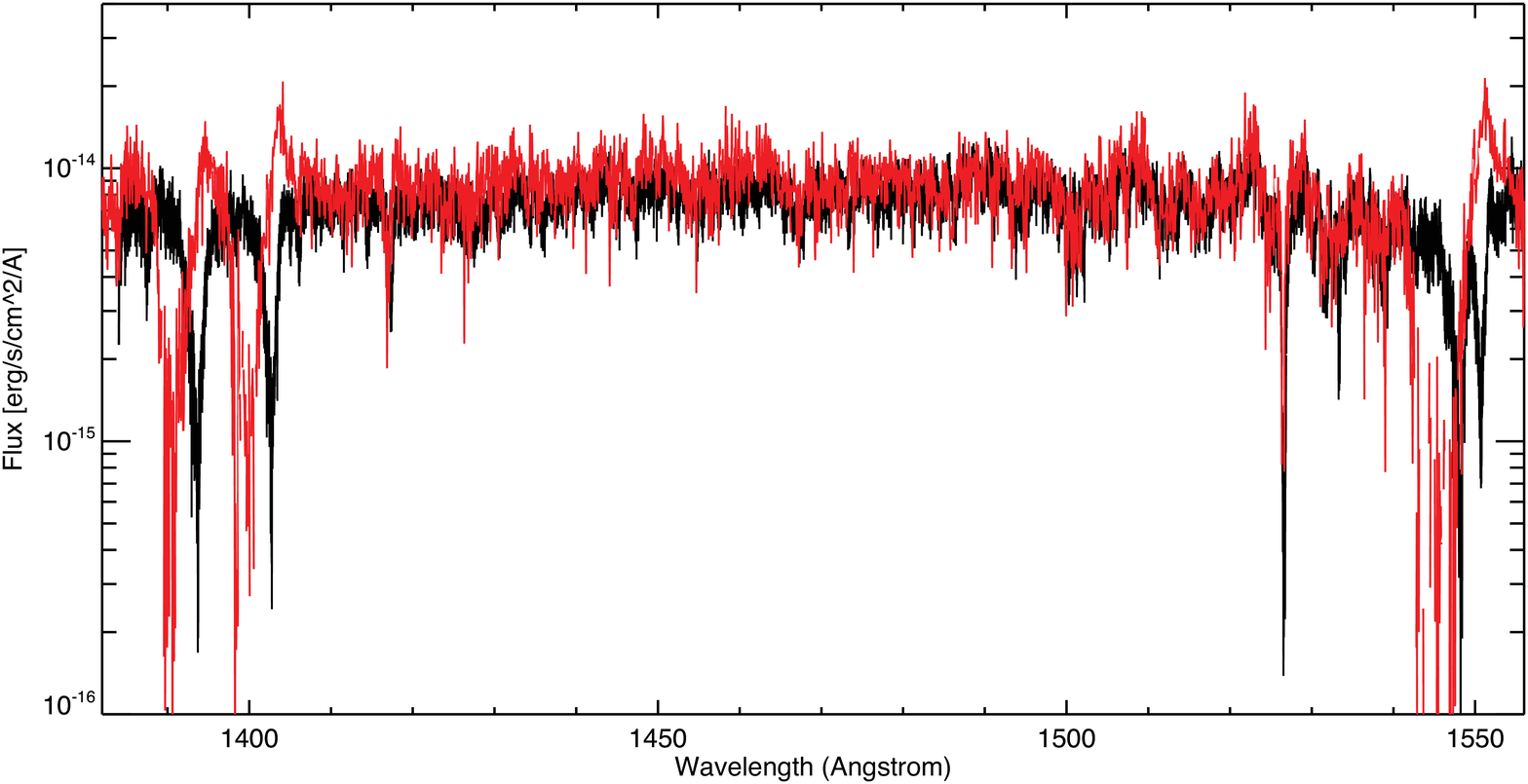}
\caption{\label{uvcomparehd} Comparison between the ultraviolet
  spectrum of SBW 1 (B1.5~Iab; black) and HD~13854 (B1~Iab; red).
  HD~13854 was scaled to roughly match the continuum flux of SBW1.
  Note the stronger \ion{Si}{iv} and \ion{C}{iv} resonance lines in
  the latter because of the higher mass-loss rate. }
\end{figure*}

In order to test whether this proposed scenario actually works for the
specific case of SBW1, we need to know the mass-loss rate of the wind
from the central star, because this was an assumed parameter in our
previous analysis \citep{smith13}.  For this reason, we proposed to
obtain UV spectra of the central star.  Optical spectra are useful for
constraining atmosphere/wind models as well, but the UV resonance
lines are usually the most sensitive probes of the wind density and
speed.  Our new observations are discussed in \S 2, our analysis of
the data including a comparison with radiative transfer models is
presented in \S 3, and a discussion of the results and implications is
given in \S 4.

\section{OBSERVATIONS}

In Cycle 20 (program GO-12924), we used the Cosmic Origins
Spectrograph (COS) onboard the {\it Hubble Space Telescope} ({\it
  HST}) to observe the UV spectrum of the BSG star at the center of
the SBW1 ring nebula.  We used FUV mode in the G160M grating with
central wavelengths of 1577 and 1600 \AA.  These two grating tilts
were combined to fill the gap between the microchannel plate detector
segments.  The COS observations were taken on 2013 March 24.  The
resulting spectral coverage was 1384$-$1777 \AA, with a spectral
resolving power $R$ of 16,000$-$18,000 (17$-$19 km s$^{-1}$).  Across
most of the spectrum the total exposure time was 11.5 ksec, although
at the edges and middle of the wavelength range (corresponding to the
COS detector gap that was filled) the effective exposure time dropped
as low as $\sim$4 ksec.  The regions with this lower signal-to-noise
ratio did not include important diagnostic lines like Si~{\sc iv} and
C~{\sc iv} discussed below.

In our analysis below, we also include a normalized visible-wavelength
spectrum of the central star.  The spectrum was obtained in 2006 with
RC Spec on the CTIO 4 m telescope, and covers roughly 3000$-$6000 \AA
\ with a spectral resolving power $R$=$\lambda$/$\Delta\lambda$ of
about 500.  This spectrum has already been published, and the
associated details of the data reduction were already presented by
\citet{sbw}.  In order to help constrain the value of the effective
gravity from the wings of Balmer lines, we also compare models to a
high-resolution echelle spectrum of SBW1's H$\alpha$ line published
previously \citep{sbw}.

\section{CMFGEN MODELS}  \label{cmfgen}

We use the radiative transfer code CMFGEN \citep{hm98} to analyze the
optical and ultraviolet spectrum of SBW1. CMFGEN self-consistently
solves the radiative transfer in a stellar atmosphere and
spherically-symmetric, stationary wind. Line and continuum formation
are calculated in the non-LTE regime. Each model is defined by the
effective temperature \teff\ (evaluated at a Rosseland optical depth
of 2/3), luminosity \lstar, effective gravity \geff, mass-loss rate
\mdot, wind terminal velocity \vinf, velocity law, and chemical
abundances. CMFGEN accounts for line blanketing, and we include the
appropriate ionization stages of H, He, C, N, O, Si, Mg, Al, Fe, and
Ni in the analysis of SBW1, with an atomic model similar to that of
\citet{crowther06} and \citet{searle08}.  CMFGEN uses an ad-hoc
velocity law as input that is typically parameterized by a beta-type
law, which is modified to smoothly match a hydrostatic structure at
high optical depths. Here we assume $\beta$=1.5.  We do not include
the effects of clumping. We refer the reader to these aforementioned
papers and \citet{hm98} for further details on CMFGEN.

We employed standard spectroscopic criteria \citep{crowther06} for
determining the stellar and wind parameters of
SBW1. Table~\ref{params} presents the inferred stellar and wind
parameters, while Figs.~\ref{modeloptical} and \ref{modeluv} display
the best fit CMFGEN model compared to the observations of SBW1 in the
optical and ultraviolet, respectively. We find $\log \geff = 2.6 \pm
0.2$ cm s$^{-2}$ based on the wings of H$\gamma$ and H$\delta$ in the
low-resolution spectrum.  Comparing to the wings of H$\alpha$ in an
echelle spectrum gives a somewhat higher value of 2.7 to 3.0 cm
s$^{-2}$ for $\log \geff$.  Our models indicate $\teff=21,000
\pm1000~\K$ based on the ionization balance of Si, using the relative
strengths of \ion{Si}{iv} $\lambda4088$, \ion{Si}{iii}
$\lambda\lambda\lambda4552, 4668, 4575$, and \ion{Si}{ii}
$\lambda4128$. The effective temperature is further supported by the
He and Fe ionization balances, using \ion{He}{ii}$\lambda~1640$ in the
ultraviolet and optical \ion{He}{i} line triplets (since singlets are
model dependent; \citealt{najarro06}), and ultraviolet \ion{Fe}{iii}
and \ion{Fe}{iv} lines.

The luminosity of SBW1 is more difficult to constrain because the
distance and reddening are uncertain.  With the CMFGEN model, we can
constrain it using the observed $V$ and $R$ magnitudes \citep{sbw},
and the flux-calibrated COS spectrum in the UV, combined with an
assumed distance. We compare the observed spectral energy distribution
with CMFGEN models computed with different luminosities, and reddened
using the parameterization from \citet{fitzpatrick99}.  The distance
to SBW1 is uncertain, derived previously from its radial velocity as
$\sim$7~kpc \citep{sbw}. We find a bolometric luminosity of
$\lstar=2.5 \pm 0.5 \times 10^4~\lsun$ (D / 7 kpc)$^2$ and a color
excess of $E(B-V)$=0.95 mag, assuming a selective-to-total extinction
parameter of $R_\mathrm{V}=3.1$.  Then again, a value of
$R_\mathrm{V}=4.8$ has been determined for clouds within the Carina
Nebula \citep{smith02}, through which SBW1 is seen, and this
extinction law would raise the luminosity.  If we add $JHK_s$
photometry from \citep{smith13}, we could not find a single extinction
law that would fit all the data.  Since SBW1 is seen through Carina,
it may be that the line-of-sight extinction has multiple components
from the Carina Nebula ($R_\mathrm{V}=4.8$) and from the normal ISM
($R_\mathrm{V}=3.1$).  If we allow $R_V$ to be a free parameter and
focus on the optical and IR photometry, we find a best fit for
$E(B-V)$=0.98 mag and $R_V$=3.8.  For these, we find a higher
luminosity of (5-6.5)$\times$10$^4$ $\lsun$.  The luminosity and mass
of the star are discussed further in Section 4.1.

Using the relative strength of optical \ion{He}{i} to H lines, our
CMFGEN analysis indicates that the He abundance of SBW1 is around
solar, with He/H=0.1 (by number). We use \ion{C}{ii} $\lambda4267$ as
diagnostic for the C abundance, \ion{O}{ii} $\lambda\lambda4070, 4317,
4367, 4596, 4650, 4661$ for O, and \ion{N}{ii} $\lambda\lambda3995,
4447, 4630$ for N. We find $12+ \log\mathrm{C/H}$=7.84, $12+
\log\mathrm{N/H}$=8.13, and $12+ \log\mathrm{O/H}$=8.30, implying that
N is enriched in comparison to C and O. The abundances of SBW1 are
within the range of typical values inferred for BSGs by
\citet{crowther06} and \citet{searle08}.

Remarkably, the observed spectrum of SBW1 does not show any clear
signature of wind emission in the ultraviolet and optical regions,
which suggests a very low value of \mdot. We computed CMFGEN models
with \mdot \ values between $10^{-10}-10^{-5}~\msun$ yr$^{-1}$. While
H$\alpha$ is contaminated by nebular emission, significant emission
would still be detectable for $\mdot \gtrsim 10^{-6}~\msun$ yr$^{-1}$,
and wind signatures could possibly be detected down to $\mdot \gtrsim
10^{-7}~\msun$ yr$^{-1}$. The ultraviolet resonance lines provide more
stringent constraints on \mdot.  Even a model with $10^{-10}~\msun$
yr$^{-1}$ still shows significant
\ion{Si}{iv}~$\lambda\lambda1394-1402$ emission, which is not detected
in the observed spectrum (upper panel of
Fig.~\ref{modeluv}). Interestingly, a CMFGEN plane-parallel model
(i.e. no wind, just photospheric emission) provides a better match to
the observed UV spectrum of the S~{\sc iv} lines (see
Fig.~\ref{modeluv}).  As such, CMFGEN models suggest that the
mass-loss rate of SBW1 is less than $10^{-10}~\msun$ yr$^{-1}$. Any
possible optically-thin wind clumping (which we have not included)
would lower this estimate even more.  As we discuss below, however,
there are reasons to suspect that some of the approximations in CMFGEN
may no longer be appropriate much below 10$^{-8}$ $M_{\odot}$
yr$^{-1}$, due to the balance of heating and cooling (and its
influence on the ionization level).  We therefore adopt this latter
value as a conservative upper limit in our analysis.  Because wind
signatures are not detected, we cannot place constraints on the wind
terminal velocity of SBW1's wind. The models analyzed here assume
$\vinf=300~\kms$ and $\beta=1.0$.

To further investigate the low value of \mdot\ for SBW1, we inspected
publicly available {\it IUE} ultraviolet spectra of BSGs. Figure
\ref{uvcomparehd} compares the UV spectrum of SBW1 to that of
HD~13854, which is a B1~Iab supergiant star \citep{searle08}. One can
clearly see that despite having a very similar photospheric spectrum,
the \ion{Si}{iv} and \ion{C}{iv} lines have P Cyg profiles and are
much stronger in HD~13854, indicating a significantly stronger wind
than in SBW1. Indeed, \citet{searle08} estimated a mass-loss rate of
$1.5\times10^{-6}~\msun$ yr$^{-1}$ for HD~13854, which is several
orders of magnitude larger than for SBW1, and consistent with
expectations.  This difference may be related to the significantly
different luminosities of these two BSGs (3.4$\times$10$^5~\lsun$ for
HD~13854), which would cause different mass-loss rates due to the
difference in radiative flux.  Even so, it is puzzling that SBW1's
wind is so weak.

\begin{table}
\caption{Stellar and wind parameters of SBW1 derived with CMFGEN.
  Some values have a range because of distance uncertainties.  As
  discussed in the text, consistency favors values of $L = 5 \times
  10^4 \lsun$, log $Q_H$ = 47.44 s$^{-1}$, and log $g_{eff}$ = 3.0.} 
\label{params}
\begin{center}
\begin{tabular}{l r}
\hline
 Quantity & Value \\
 \hline
  Luminosity \lstar\ (\lsun)		    & (2.5-6.5) $\times$ 10$^4$  \\
  Effective Temp. \teff\ (K, at $\tauross$=2/3) & 21,000 $\pm$1000 \\
  Log H ionizing photon flux $Q_H$ (s$^{-1}$) & 47.14-47.44 \\
  Log Effective Gravity (cm s$^{-2})$        & 2.6-3.0 \\
  Wind Terminal Velocity (km s$^{-1}$)       & 300 (assumed) \\
  Mass-Loss Rate ($M_{\odot}$ yr$^{-1}$)      & $< 1.0\times10^{-10}$ \\
  He/H (by number)                          &  0.1 \\
  $12+ \log\mathrm{C/H}$ (number)           &  7.84 \\
  $12+ \log\mathrm{N/H}$ (number)           &  8.13 \\
  $12+ \log\mathrm{O/H}$ (number)           &  8.30 \\  
 $\log\mathrm{N/C}-\log\mathrm{N_\odot/C_\odot}$ (number) & +0.90\\
 $\log\mathrm{N/O}-\log\mathrm{N_\odot/O_\odot}$ (number) & +0.44\\
 \hline
\end{tabular}
\end{center}
Note: Although this mass-loss rate of 10$^{-10}$ $M_{\odot}$ yr$^{-1}$
appears to be a reasonable upper limit resulting from the
CMFGEN model, in the text we discuss why a higher upper limit of
10$^{-8}$ $M_{\odot}$ yr$^{-1}$ is probably more important, due to a
possible lack of cooling in the wind.
\end{table}

\section{DISCUSSION}

\begin{figure*}
\includegraphics[width=5.5in]{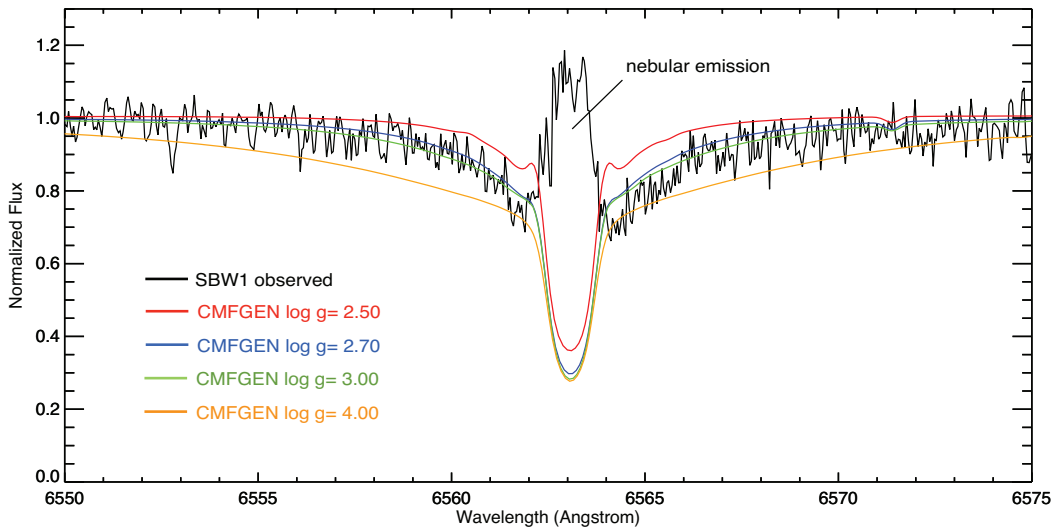}
\caption{\label{halpha} . Detail of the H$\alpha$ line profile in an
  echelle spectrum of the star obtained with EMMI (from
  \citealt{sbw}).  The emission at line center is due to nebular
  emission from the ring (this is clear in resolved long-slit spectra;
  see \citealt{sbw}), but the line wings trace the photospheric
  absorption profile due to pressure broadening. The rotation rate is
  only $\sim$40 km s$^{-1}$ \citep{taylor14}, so rotation does not
  alter the broad line wings.  To this high-resolution spectrum, we
  compare a CMFGEN model with the same parameters as discussed above,
  except that we explore different values of the effective gravity.
  From this comparison, it is evident that log $g_{\rm eff}$=2.5 (cgs
  units) is too low and 4.0 is too high, but log $g_{\rm eff}$ values
  of 2.7-3.0 provide a good match to the data, with 3.0 being somewhat
  preferred in some wavelength ranges.}
\end{figure*}

\begin{figure*}
\includegraphics[width=5.5in]{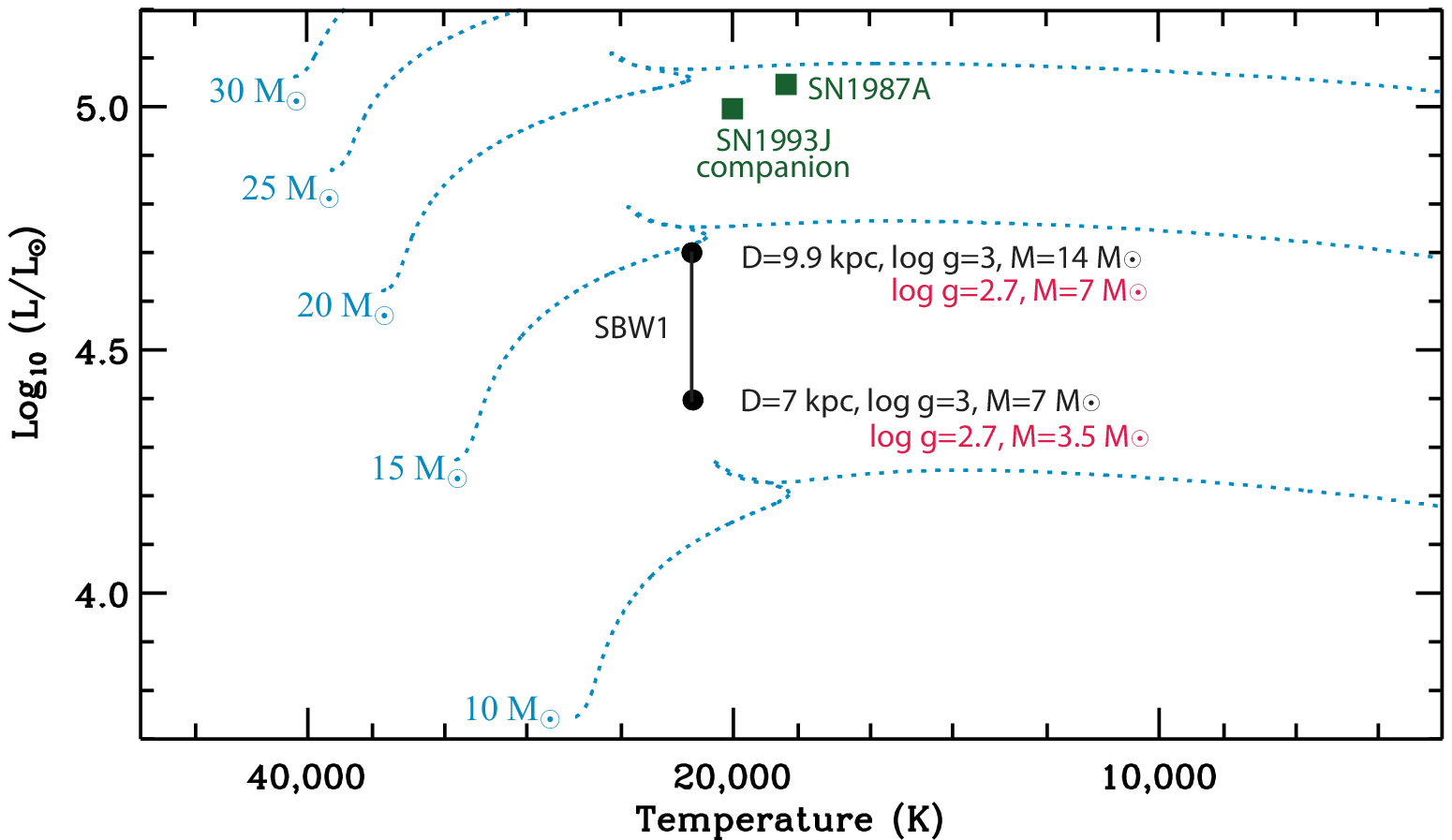}
\caption{\label{hrd} HR Diagram with representative single-star
  evolution tracks from \citet{brott11}.  We denote the location of
  the progenitor stars of SN~1987A \citep{maund+04} and the companion
  of SN~1993J \citep{fox+14}, as compared to SBW1. For SBW1, the two
  black filled dots show the luminosity indicated by our CMFGEN model
  for assumed distances of $D_7$=7 kpc and $\sqrt{2} \times D_7$=9.9
  kpc, while the red and black text give the implied present-day
  stellar masses for log $g_{\rm eff}$=2.7 and 3, respectively. The
  implication is that an assumed distance of 7 kpc is too small,
  because $g_{\rm eff}$ gives stellar masses that are far below the
  mass one would infer from comparing the luminosity to evolutionary
  tracks.  A slightly larger distance of 9.9 kpc, on the other hand,
  gives double the luminosity and stellar mass, and is in much better
  agreement with the luminosity for evolutionary models with that mass
  if we adopt log $g_{\rm eff}$=3. }
\end{figure*}

\subsection{Central star properties}

Our CMFGEN analysis confirms many of the physical paramaters that had
been inferred previously from photometry and spectral type.  The value
of $\teff$ = 21,000~K that we derive from the CMFGEN model is the same
as assumed previously from the B1.5~Iab spectral type and spectral
energy distribution \citep{sbw,smith13}.  The luminosity derived
previously from the SED was uncertain (0.5-1)$\times$10$^5$ $\lsun$.
Our CMFGEN model gives a somewhat lower value of 2.5$\times$10$^4$
$\lsun$ $D_7^2$, where $D_7$ is the distance relative to our assumed
value of 7 kpc.  As noted earlier, however, the true luminosity would
be raised to (5-6.5)$\times$10$^4$ $L_{\odot}$ if we adopted a larger
value of $R_V$, as may be appropriate.  Below, we also find that the
likely distance is larger than 7 kpc.

Interestingly, with the weak wind of SBW1, we can use the effective
gravity $g_{\rm eff}$ from the model to place constraints on both the
true luminosity and present-day mass if we assume that the
spectroscopically derived mass $M_{\rm spec}$ is comparable to the
evolutionary mass $M_{\rm evo}$.  \citet{lk14} have discussed this in
detail, and concluded that $M_{\rm spec}$ is usually a reliable
representation of the true stellar mass as long as the star is not
close to the Eddington limit (i.e. for moderately massive and
intermediate-mass stars).  Since SBW1 has no detectable signatures of
a wind, we surmize that it is nowhere near its Eddington limit.  Our
model derived from a comparison to low-resolution spectra gave log
$g_{\rm eff}$=2.6 (Table 1).  However, by comparing CMFGEN models to
higher resolution spectra of Balmer lines, as shown in
Figure~\ref{halpha}, we favor a somewhat higher value of 2.7-3.0 for
log $g_{\rm eff}$.  This difference arizes because the
lower-resolution spectra are compromised by nebular emission that
affects the line profiles, whereas the stellar Balmer line wing shapes
are resolved in the echelle spectrum. From the definition of $g_{\rm
  eff} = GM/R^2$ we can write

\begin{displaymath}
  M_{\rm spec} = \frac{g_{\rm eff} \, R_*^2}{G} = \frac{g_{\rm eff} \, L_*}{4 \pi \sigma \, G \, T_{\rm eff}^4}
\end{displaymath}

\noindent where $M_{\rm spec}$ is the present-day spectroscopically
derived stellar mass, $R_*$ and $L_*$ are the star's photospheric
radius and bolometric luminosity, respectively, $\sigma$ is the
Stefan-Boltzmann constant, and G is the gravitational constant.  We
note that the uncertainty is dominated by errors in $g_{\rm eff}$
rather than errors in $T_{\rm eff}$. Inserting fiducial values of
$T_{\rm eff}$ = 21,000 K, $g_3 = (g_{\rm eff} / 10^{3} {\rm cm \,
  s^{-2}})$ and $L_* = 2.5 \times 10^4 L_{\odot}$ ($D_7$)$^2$, where
$D_7$ is the adopted distance relative to 7 kpc, we then have

\begin{displaymath}
  M_{\rm spec} \, \approx \, 7 \, M_{\odot} \, ( g_3 \, D_7^2) 
\end{displaymath}

\noindent for the {\it present-day} stellar mass as indicated by the
effective gravity.  We can then attempt to constrain the actual
luminosity and mass of SBW1 by seeing what combinations of $D$, $L_*$,
$g_{\rm eff}$, and $M_{\rm spec}$ give values consistent with
evolutionary models.  Figure~\ref{hrd} shows a Hertzsprung-Russel (HR)
diagram comparing the inferred luminosity of SBW1 to single-star
evolutionary tracks \citep{brott11}, for reference.  This comparison
shows that the lower luminosity for an assumed distance of 7 kpc,
combined with the $g_{\rm eff}$ indicated by the spectrum, gives a
stellar mass of $\sim$7 $M_{\odot}$ --- but this is much lower than
one expects in this region of the HR Diagram.  For a slightly larger
distance of 9.9 kpc, however, the luminosity and $M{\rm spec}$ rise by
a factor of two, and importantly, are then in very good agreement with
the expected luminosity for a 14-15 $M_{\odot}$ evolved star.  In
fact, the $T_{\rm eff}$ and $L_*$ we derive would agree very well with
the hook in the evolutionary track for a 14 $M_{\odot}$ star that
occurs after core H exhaustion, which would seem to make sense with
the blue supergiant spectral type.  Of course, the comparison to
single-star evolutionary tracks may be misleading if SBW1 is the
result of binary evolution that may alter its $L$/$M$ ratio; one could
argue that it would be appropriate to compare SBW1's spectroscopic
mass to models for a merger product or mass gainer.  Such a comparison
might favor a slightly different combination of $D$, $g_{\rm eff}$,
$L_*$, and $M_*$, especially if these values evolve in the
$\sim$10$^4$ yr after a merger or mass transfer episode.

We could, of course, also achieve a luminosity of 5$\times$10$^4$
$L_{\odot}$ with a smaller distance (8 kpc, say) and a slightly higher
$R_V$ value.  The luminosity is unlikely to be much lower than
2.5$\times$10$^4$ $\lsun$, however, due to the fact that the optical
spectrum has a supergiant luminosity class, and that the implied mass
from $g_{\rm eff}$ would be too low for the corresponding $L$.  Based
on this comparison, we therefore favor values of log $g_{\rm eff}$=3.0
cm s$^{-2}$ and $L_*$ = 5$\times$10$^4$ $L_{\odot}$ for SBW1.

Altogether, these parameters make the central star of SBW1 only a
little hotter than Sk$-$69$^{\circ}$202, and about 50\% of its
bolometric luminosity.  It has an effective initial mass of around
14~$\msun$, as compared to 18~$\msun$ for Sk$-$69$^{\circ}$202
\citep{arnett89,arnett+89}.  From $L \propto R^2 T_{\rm eff}^4$, the
implied stellar radius is of order 15-20 $\rsun$, and so the surface
escape velocity is about the same as for the Sun or slightly lower.

The chemical abundances we derive from the photospheric spectrum show
basically Solar composition, except for an enhanced N abundance that
is elevated by a factor of 3 or 8 compared to Solar N/O or N/C ratios,
respectively.  This, too, is quite similar to the enhanced N
abundances inferred from the emission-line spectrum of the ring around
SN~1987A, and is indicative of significant CNO processing present at
the star's surface.

Will the central star of SBW1 be the next Galactic SN?  The dynamical
age of the nebula is about 10$^4$ yr, similar to SN~1987A, so perhaps
it is a good candidate.  Of course, the uncertainty of such a clock is
huge.  Aside from the nebular age and an analogy to SN~1987A, we have
little from which to infer the time until the impending core collapse.

\begin{figure*}
\includegraphics[width=4.5in]{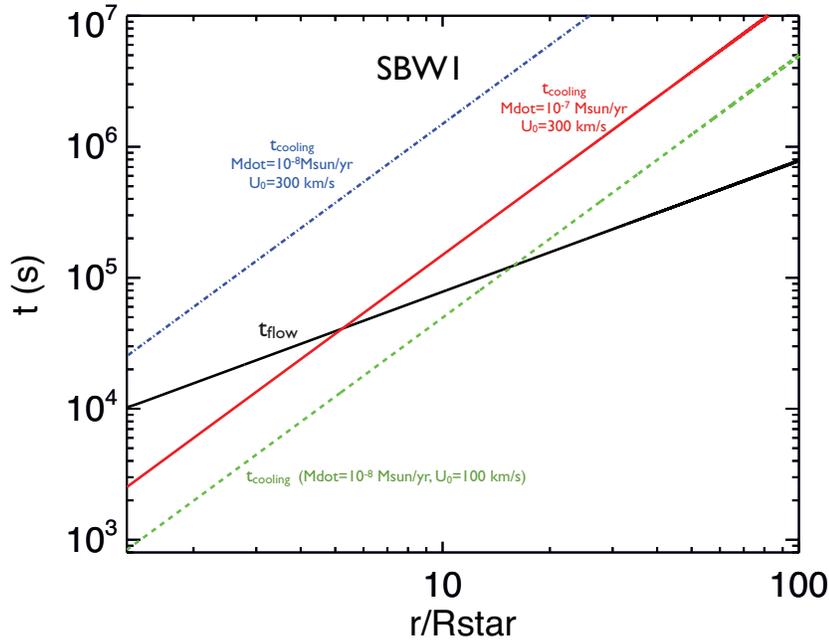}
\caption{\label{flow}Plot of the flow timescale (black solid) compared
  to the cooling timescale in a CMFGEN model as a function of radius
  for some representative assumed values of mass-loss rate and shock
  velocity $U_0$.  If the wind outflow speed is 300-500 km s$^{-1}$,
  then the typical speed of internal shocks in the wind is probably
  100--200 km$^{-1}$ or less, and very likely less than 300 km
  s$^{-1}$.  Thus, we see that for expected shock speeds, the wind
  cooling timescale becomes comparable to the flow timescale in the
  inner wind when the mass-loss rate drops below 10$^{-8}$ $M_{\odot}$
  yr$^{-1}$.  As such, it is possible that UV diagnostics become less
  reliable at such low wind densities due to increased ionization.}
\end{figure*}

\subsection{Stellar wind properties}

The most significant observational result from our COS spectrum is the
lack of any strong wind features in the spectrum, which is very
unusual for a blue supergiant.  Consequently, the most interesting
result from our quantitative CMFGEN analysis is the astonishingly low
derived mass-loss rate of SBW1.  Comparing our lowest mass-loss rate
CMFGEN models (which still show some evidence of wind emission) to the
observations, which show none, implies and upper limit of $\dot{M} <
10^{-10}$ \ $\msun$ yr$^{-1}$ for an assumed terminal wind speed of
300 $\kms$.  However, this assumes that the UV resonance lines are
modeled correctly at such low wind densities.

A cautionary remark relates to the so-called ``weak-wind problem''
(see \citealt{smith14} for a review), where UV-diagnostics of
late-type O dwarfs yield mass-loss rates that are 100 times lower than
expected from the Vink et al.\ recipe and from H$\alpha$
diagnostics. The cautionary comment is that an independent method of
deriving the mass-loss rate based on the structure of a bow shock
around $\zeta$~Oph gives a mass-loss rate estimate that is $\sim$10
times higher than UV diagnostics, but still an order of magnitude
lower than expected from standard mass-loss prescriptions
\citep{gvaramadze12}.  Thus, there are some indications that the
weak-wind problem for late O dwarfs is perhaps not as severe as
indicated by UV estimates.  Thus, when mass-loss rates are low, CMFGEN
and similar models might underestimate the mass-loss rate somewhat
based on UV diagnostics.  This may be caused by inefficient cooling at
low wind densities, so that shocks within the wind keep the ionization
level higher than expected \citep{bouret15,puebla16}.  Does some
version of this weak wind problem translate to BSG winds such as the
case of SBW1?

Figure~\ref{flow} shows how the flow timescale in the wind compares to
the cooling timescale for some representative assumed values of the
mass-loss rate and the collision speed $U_0$ of internal shocks in the
wind.  The temperature and ionization balance of the wind depends on
heating by shocks within the flow, and cooling, which depends on the
density.  If the cooling timescale becomes long at low densities, the
wind may expand before it can cool, and so the ionization in the inner
wind may be higher than in a CMFGEN model.  Typical wind speeds for an
early B supergiant would be around 500 km s$^{-1}$, and we would
expect shocks within the clumpy outflowing wind to be some fraction of
that -- perhaps 100-200 km s$^{-1}$ and almost certainly less than 300
km s$^{-1}$ (a caveat is that we can't be certain about the value of
the wind speed, since we don't actually detect any wind absorption;
thus, it remains possible that SBW1's wind might be faster than
typical winds for B1.5 supergiants, which might raise the allowed
mass-loss rate).  Therefore, in Figure~\ref{flow} we should expect
SBW1's wind to reside somewhere above the green dashed line and below
the blue dash-dotted line (for $U_0$=100 and 300 km s$^{-1}$,
respectively).  For an intermediate shock speed around 200 km
s$^{-1}$, for example, a mass-loss rate below 10$^{-8}$ $M_{\odot}$
yr$^{-1}$ would make the cooling timescale and the flow timescale
about the same in the inner wind (a few stellar radii).  This means
that below 10$^{-8}$ $\msun$ yr$^{-1}$, the bulk of the wind might
remain hotter than in the CMFGEN model, and would be harder to observe
in the typical UV diagnostic lines that we are referrring to.  Thus,
if the mass-loss rate drops much below 10$^{-8}$ $M_{\odot}$
yr$^{-1}$, we cannot be confident that CMFGEN is properly treating the
relevant physics, whereas above this, we should begin to see some
evidence of a wind in the UV lines.  We adopt 10$^{-8}$ $M_{\odot}$
yr$^{-1}$ as a fairly conservative upper limit to the mass-loss rate
of SBW1, rather than 10$^{-10}$ $M_{\odot}$ yr$^{-1}$, due to this
uncertain treatment of the cooling and ionization balance in CMFGEN at
such low mass-loss rates.

Our hypothesis that the wind of SBW1 has too low a density to cool ---
and therefore remains hot --- can be tested.  This hypothesis would
predict detectable X-ray, EUV, and possibly FUV emission signatures
from the wind, which may be verified with future observations. An
observational determination of $L_X/L_{\rm Bol}$ with future X-ray
observations would thus help provide a direct constraint on the
mass-loss rate of SBW1 and the amount of shock heating within the
wind.  At the stellar temperatures appropriate for SBW1's spectral
type, CMFGEN does not predict any N~{\sc v} or O~{\sc vi} features,
but CMFGEN's reatment of the wind is not appropriate if the wind
remains hot as we suspect.  If low density inhibits cooling, we can
make a qualitative prediction that N~{\sc v} and O~{\sc vi} may be
observed (see, e.g., \citealt{bouret15,puebla16,zsargo08}), but a more
detailed model beyond CMFGEN's current capabilities would be needed to
derive a specific line strength for a quantitative mass-loss rate.

Even this revised upper limit to the mass-loss rate of 10$^{-8}$ is
much lower than one would expect for this star.  For example, from the
mass-loss prescriptions given by \citet{vink01}, we would expect
$\dot{M}$ = 1.2$\times$10$^{-7}$ $\msun$ yr$^{-1}$ for $L = 5 \times
10^4 \lsun$, for line-driven winds at the appropriate $\teff$ of SBW1
(note that 21,000 K places this on the cool side of the bistability
jump).  Our observationally derived upper limit to the mass-loss rate
for SBW1 is more than 10 times lower than this expected value, even
with no reduction in the observed value to correct for clumping.

Why is the wind of SBW1 so weak as compared to expectations, and as
compared to other observed BSGs?  The solution to this puzzle may hold
important clues related to the origin of the ring nebula and the
star's evolutionary history, and perhaps also for extragalactic SNe
that appear similar to SN~1987A.

For SN~1987A, the physical properties of the pre-SN stellar wind were
uncertain, but some considerations also pointed to an anomalously low
mass-loss rate compared to other BSGs.  On the one hand, models
derived from interpreting the early radio observations in the context
of free-free self absorption of the SN radio emission by the freely
expanding wind yielded a relatively high mass-loss rate of order
3.5--6 $\times$ 10$^{-6}$ $M_{\odot}$ yr$^{-1}$ with $v_w$=550 km
s$^{-1}$ \citep{cf87,lf91,cd95}.  On the other hand, hydrodynamic
interacting-winds models used to explain the formation of the nebula
required much weaker BSG winds in order to reproduce the slow
expansion speed of the equatorial ring \citep{bl93,ma95}.  To keep the
ring expanding at the slow observed value of $\sim$10 km s$^{-1}$,
these models would require upper limits to the mass-loss rate and wind
speed of $\dot{M} < 3 \times 10^{-7}$ $M_{\odot}$ yr$^{-1}$ and $v_w <
300$ km s$^{-1}$.  \citet{bl93} suggested that this discrepancy might
be explained if the star's mass-loss rate increased in the last
decades or century leading up to the moment of explosion, but
\citet{cd95} suggested that synchrotron self-absorption, rather than
free-free self absorption by the wind, might explain the early radio
observations.

Further indication that the progenitor star's mass-loss rate was low
compared to normal BSGs came from the rebrightening in the radio at
$\sim$1500 days after the SN
\citep{ss92,ss93,gaensler97,gaensler00,manchester02,zanardo13}.  This
rebrightening was attributed to the collision between the fast SN
ejecta and an H~{\sc ii} region from the photoionized RSG wind, as
noted in the Introduction \citep{cd95}.  Assuming that the interior
region was filled with a relatively low-density freely expanding BSG
wind, \citet{cd95} showed that this collision could occur at the
observationally inferred radius of $\sim$0.1 pc with a model that
adopted $\dot{M} = 7.5 \times 10^{-8}$ $M_{\odot}$ yr$^{-1}$ and $v_w
= 450$ km s$^{-1}$. Later models refined this value, in some cases
including constraints from the evolution of X-ray emission, to even
lower values of around 5$\times$10$^{-9}$ $M_{\odot}$ yr$^{-1}$ or
less \citep{vikram07,dewey}.

This is a very low mass-loss rate for a BSG star of $\sim$10$^5$
$L_{\odot}$. According to the standard recipie for hot star mass-loss
rates usually used in evolutionary codes \citep{vink01}, a star with
log($L/L_{\odot}$)=5, $T_{\rm eff}$=21,000~K, and $M$=18 $M_{\odot}$
should have a mass-loss rate of 4.8$\times$10$^{-7}$ $M_{\odot}$
yr$^{-1}$ at LMC metallicity.  The mass-loss rate inferred for the
progenitor of SN~1987A based on the expansion of the blast wave is at
least 6 and as much as 100 times lower than this expected value. This
appears very similar to the case of SBW1 outlined above.

Thus, both the progenitor of SN~1987A and SBW1 seem to share the
pecularity that they have BSG winds that are extremely weak compared
to the expected wind strength for their stellar parameters.  This is
not the case for the other two well-studied Galactic analogs with ring
nebulae; both Sher~25 and HD~168625 have strong H$\alpha$ wind
emission and have mass-loss rates that are normal (Sher~25) or strong
(HD~168625) compared to other BSGs \citep{smartt02,nota96}.  Although
accounting for clumping has been argued to require a reduction to
mass-loss rate recipies by factors of 3-5 (see \citealt{smith14} for a
review), the deficits for SN~1987A and SBW1 are greater than this (and
again, we did not include a clumping correction for SBW1).

In models that aim to explain the formation of SN~1987A's triple ring
nebula with a merger \citep{mp07,mp09}, BSG mass-loss rates of (1-2)
$\times$ 10$^{-7}$ $M_{\odot}$ yr$^{-1}$ are adopted to shape the ring
nebula (e.g., Table 4 in \citealt{mp09}).  Recently, \citet{orlando15}
adopted this same value for the mass-loss rate in their simulations of
the SN interaction with the CSM, although they did not explore the
impact of other assumed values for the mass-loss rate.  These are
higher than the observationally inferred values for SN~1987A (from the
time history of radio emission, as noted above) and for SBW1.  It is
therefore unclear if interacting stellar winds can provide a viable
physical explanation for the shaping of the nebulae around SN~1987A
and SBW1.  The issue of pressure balance is discussed more below.

\subsection{Implications for the nebula and the pre-SN evolution of
  SN~1987A}

Previous studies have discussed the formation of bipolar and ring
nebulae, like the ones around SN~1987A and SBW1, in the context of
interacting winds where a fast BSG wind expands into a slower and
asymmetric RSG wind with an equatorial density enhancement (see the
Introduction).  However, a somewhat different scenario was discussed
wherein a fast BSG interacts with an H~{\sc ii} region or
photoevaporative flow for the specific cases of SN~1987A \citep{cd95}
and SBW1 \citep{smith13} based on the inferred density structure
inside the ring, which is inconsistent with a simple interacting winds
scenario.  In this section, we discuss how the extreme weakness of the
BSG wind from SBW1 requires further modification to the story. 

In this scenario, the location of the shock between the BSG wind and
the ionized photoevaporative flow is determined by the mass-loss rate
and speed of the wind from the central star, balanced by the pressure
of the photoevaporative flow.  Specifically, ram pressure of the
stellar wind $\rho \, v^2$ is balanced by the thermal pressure of the
ionized gas inside the ring nebula.  The photoevaporation rate of the
ring that is the source of gas and dust in the H~{\sc ii} region
depends on geometry and the ionizing photon flux of the star, $Q_H$,
which is given in Table 1.  However, in this case we can avoid the
uncertainty introduced by the detailed geometry of the ring (clump
size, ring height, whether gas in the walls of an hourglass
contributes, etc.)  because spectral observations of the nebula (the
H$\alpha$ emission measure and the [S~{\sc ii}]
$\lambda\lambda$6717,6731 line instensity ratio in the spatially
resolved diffuse interior of the ring) directly constrain the density
of the ionized flow filling the inside the ring to be roughly 300-500
cm$^{-3}$ \citep{smith13}. Thus, the ionized gas pressure is directly
constrained observationally, and so the pressure there is known
regardless of the geometry that creates it.  While the pressure within
this H~{\sc ii} region is roughly uniform, the ram pressure of the
wind drops with radius from the star if we assume a steady BSG wind
($R^{-2}$ density profile). Then $R$ is the radius where the two
balance, given by

\begin{displaymath}
  R \ = \ 0.05 \ \Big{(} \dot{M}_{-7} \ V_{300} \Big{)}^{1/2} \Big{(}\frac{n_e}{500 \ {\rm cm}^{-3}} \Big{)}^{-1/2} \ {\rm pc},
\end{displaymath}

\noindent 
where we have assumed $T = 10^4$\,K in the H~{\sc ii} region,
$\dot{M}_{-7}$ is the BSG wind mass-loss rate in units of
$10^{-7}$\,M$_{\odot}$\,yr$^{-1}$, and $V_{300}$ is the wind speed in
units of 300\,km\,s$^{-1}$.  We assumed a value for $V_{\rm BSG}$ of
300\,km\,s$^{-1}$, as above.  These fiducial values are similar to the
values adopted for the progenitor of SN~1987A by \citet{cd95}.

With these values, the stand-off shock will be at $R \approx 0.05$\,pc
from the BSG.  This is about 25\% of the radius of the ring (note that
both SN~1987A and SBW1 have the same ring radius of $\sim$0.2 pc).  In
the case of SBW1, 25\% of the ring radius roughly matches the location
of the observed inner peaks of hot dust and enhanced H$\alpha$
emission in images, which is why we chose these fiducial values.
Since the innermost dust near the shock front will be the hottest and
brightest because it is radiatively heated by the star, we argued
\citep{smith13} that this physical scenario may give a plausible
explanation for the structures inside the ring.  We subsequently
proposed to obtain UV spectra to directly constrain SBW1's mass-loss
rate in order to test this picture.

We were therefore surprised to find a mass-loss rate for SBW1 that is
at least an order of magnitude lower than the fiducial value above.
With $\mdot <$ 10$^{-8}$ $\msun$ yr$^{-1}$ (a conservative upper
limit), the radius of the stand-off shock between the BSG wind and the
ionized photoevaporative flow should be much smaller, roughly $<$0.015
pc or only about 5--10\% of the ring's radius.  Essentially, the BSG
wind is so weak that it would be overwhelmed by the gas pressue of the
photionized photoevaporative flow.  Colliding winds may therefore have
difficulty explaining the pile-up of dust at the location of the
observed IR peaks in images \citep{smith13}.  A renewed investigation
of this problem using hydrodynamic simulations is warranted.




What, then, causes the peaks of dust emission at $\sim$25\% of the
radius of the ring (at $R$$\approx$0.05 pc from the star)?  As noted
in our previous paper \citep{smith13}, the observed dust temperature
estimated from the SED is only about 190~K (and the expected
equilibrium temperature is even lower at that radius) so 0.05 pc
cannot mark the dust vaporization radius.  Something else must hold
back the dust and prevent it from flowing closer to the star.  A
possibility is that direct stellar radiation pressure on dust grains
helps keep them at bay, and that collisions couple this radiation
pressure on dust to the gas.  Indeed, the magnitude of the radiation
pressure $L / (4 \pi R^2 c)$ inside the ring, for our derived stellar
parameters of SBW1, is comparable to or greater than the inferred
ionized gas pressure for $T = 10^4$ K and $n_e$=500 cm$^{-3}$,
suggesting that direct radiation pressure on dust should affect the
structure and dynamics of the interior of the ring.

So far, radiation pressure has not been included in simulations aiming
to explain the origin and shaping of BSG rings like the ones around
SN~1987A and SBW1.  However, the weakness of the observed wind from
SBW1 reported here (as well as the inferred weakness of the wind of
SN~1987A's progenitor) suggest that this should be undertaken.
Examining the hydrodynamics including radiation pressure is beyond the
scope of this paper, but we note that the problem is reminiscent of
recent studies of the dynamics and structure of dusty H~{\sc ii}
regions, where radiation pressure on dust is also found to be
important \citep{km09,draine11,kim+16}.  The relative influence of
radiation pressure is even stronger in the case of SBW1 due to its
extremely weak stellar wind for its luminosity.

Another possibility, which is difficult to rule out, is that the inner
dust peaks arise from a past eruptive mass ejection akin to LBV
eruptions \citep{smith11}.  While the BSG wind cannot form dust in its
steady wind, it could potentially form dust in an episodic ejection of
a dense shell (see, e.g., \citealt{kochanek11}).  This dust shell
might then expand until it is stopped by the pressure of the
photoevaporative flow, leaving a cavity in its wake to be filled by
the very weak BSG wind.  In this case it would be the momentum of the
(hypothetical) eruptive mass ejection rather than the ram pressure of
the BSG wind that would set the location of the inner dust peaks.
This scenario is admittedly somewhat {\it ad hoc}, but there is
precedent for it.  Sequential episodic ejections of rings have been
inferred based on direct proper motions of the ring nebula around the
massive binary RY Scuti \citep{smith+11}, for example.  For a somewhat
different type of system, hydrodynamic simulations of nova eruptions
inside a slow, equatorially concentrated CSM produced by RLOF can
yield a similar torus structure with inner density peaks
\citep{booth16}.

\subsection{Spindown}

The very weak wind of SBW1 has an important consequence regarding the
star's rotational evolution (e.g., \citealt{meynet11}).  Such a low
mass-loss rate will impair the star's ability to shed angular momentum
via its wind.  SBW1 currently has a rather slow rotation rate, with an
equatorial rotation speed of only about 40 km s$^{-1}$
\citep{taylor14}, which is only about 5\% of its critical rotation
speed.

The current slow rotation rate coupled with the currently observed
very low wind mass-loss rate presents a puzzle in connection with the
observed ring nebula.  As noted in the Introduction, most scenarios to
explain the existence of ring nebulae like the ones around SN~1987A
and SBW1 invoke either (1) mass transfer through RLOF in an
interacting binary (which would spin up the mass gainer and then shed
mass through the outer Lagrange point), (2) the merger of a close
binary system resulting in a rapid rotator that excretes a disk, ring,
or torus in the merger, or (3) post-RSG contraction to a BSG, spinning
the star up to a rapidly rotating star that sheds an equatorial disk.
All of these include a star that is rotating at or close to critical
rotation when the ring is ejected.  In the case of SBW1, the ring is
only about 10$^4$ yr old \citep{sbw,smith13}.

The puzzle, then, is how a star can go from (presumably) nearly
critical rotation (several 10$^2$ $\kms$) to being such a slow rotator
(only 40 $\kms$) in such a short time if its wind is very weak.  In
this time, the star would shed a tiny fraction ($\sim$10$^{-5}$ or
less) of its total mass.

Magnetic breaking would be the key mechanism to spin down the star,
and indeed, it has been suggested that a stellar merger event - which
might eject a ring - might also lead to very strong stellar magnetic
fields \citep{schneider16}.  However, one expects the loss of angular
momentum via magnetic breaking to be directly proportional to the
mass-loss rate, which in the case of SBW1 is exceedingly low.  Even
for massive stars with very strong (a few to several kG) fields and
stronger winds, the spin-down timescale is a few to several Myr
\citep{uddoula09}, not 10$^4$ yr.  Indeed, using a parameterized
estimate for the spin-down time from Equation 25 in \citet{uddoula09},
and adopting a generous 3 kG magnetic field, the parameters we
estimate for SBW1 would suggest a spin-down timescale of $>$6 Myr.  It
is therefore difficult to understand how the star could have slowed
its rotation rate during the age of the nebula of only 10$^4$ yr
unless the mass-loss rate was much higher in the past.  

Ways out of this puzzle may require some different ideas.
Observationally, at least, the gas and dust that partly fills the
interior of the ring \citep{smith13} could be interpreted as evidence
for a previous high $\mdot$ \ phase, with a slow, dense, dusty wind
that followed a merger and the ring's ejection.  Perhaps a highly time
variable wind or eruption needs to be invoked to help resolve this
issue.  Alternatively, perhaps a merger scenario different from
proposed models is in order.  For example, a merger of two blue stars
(rather than a RSG) may lead to an envelope that is out of thermal
equilibrium, as rotational energy is used to heat the merger product's
envelope.  The subsequent inflation of that envelope might allow a
merger product to have a slow surface rotation rate at such a young
age after a merger event.  It is difficult to see how very rapid
rotation can be avoided in a scenario wherein a merger occurs as a
RSG, and then the merger product contracts to the blue while also
maintaining a very low mass-loss rate.  The low BSG wind mass-loss
rate that we derive here is therefore an important constraint for
models that aim to explain the origin and shaping of such ring nebulae
with a merger event. 

SBW1 may be an interesting target for spectropolarimetry to
investigate the possibility of a strong magnetic field, although this
may be complicated by large interstellar polarization.  It is
interesting to note that some models predict that magnetic massive
stars can avoid the RSG phase altogether, staying blue and exploding
as BSGs \citep{petermann15}.  How this BSG star can avoid driving a
much stronger wind with its current luminosity remains puzzling.

\section{SUMMARY}

We obtained the UV spectrum of the blue supergiant SBW1 using {\it
  HST}/COS, with the aim of measuring the star's mass-loss rate in
order to test a hypothesis regarding the shaping of the ring nebula.
A CMFGEN model was used to analyze this spectrum.

We were surprised to find that the UV spectrum showed no signatures of
wind emission or absorption, and the CMFGEN model yielded a
conservative upper limit to the mass-loss rate of 10$^{-10}$
$M_{\odot}$ yr$^{-1}$.  However, we suspect that the mass-loss rate of
SBW1 is low enough that the UV diagnostics modeled by CMFGEN are not
good tracers of the wind, probably because the wind is unable to cool
at such low density.  We find that the cooling timescale is similar to
or longer than the flow timecale in the inner wind if the mass-loss
rate falls below 10$^{-8}$ $M_{\odot}$ yr$^{-1}$, and adopt this value
as a more likely upper limit for the wind mass-loss rate.

Even 10$^{-8}$ $M_{\odot}$ yr$^{-1}$ is much lower than expected for a
BSG with SBW1's physical parameters.  This may present a problem for
shaping the ring nebula with stellar wind interaction alone.  We therefore
speculate that radiation pressure on dust entrained in the
photoevaporative flow off the ring may play an important dynamic role
in shaping the nebula.  Moreover, the very weak wind will inhibit the
star's ability to shed angular momentum, which is problematic given
SBW1's slow observed rotation speed of $\sim$40 km s$^{-1}$.  Even
with a generous magnetic field, we find that the likely spin-down
timescale is several Myr, which is much longer than the
$\sim$10$^4$ yr age of the ring nebula.  This makes it difficult to
understand how the ring could have been ejected in a merger event,
which would be expected to leave behind a rapidly rotating star.

Based on the time dependence of radio emission, SN~1987A was also
inferred to have a very weak wind for it's progenitor's physical
parameters, so our finding for SBW1 may impact ideas about SN~1987A's
pre-SN evolution and the shaping of its nebula.

\section*{Acknowledgements}

\scriptsize 

We thank John Hillier for maintaining CMFGEN and for continuing to
make it available to the community, and also for advice on
interpreting the very low mass-loss rate implied by the model in this
case.  Support was provided by the National Aeronautics and Space
Administration (NASA) through HST grants GO-12924, GO-11637, GO-11977,
GO-13390, and GO-13791 from the Space Telescope Science Institute,
which is operated by AURA, Inc., under NASA contract NAS5-26555.
N.S.\ also received partial support from NSF grants AST-1210599 and
AST-1515559.  J.H.G.\ was supported by an Ambizione Fellowship of the
Swiss National Science Foundation.

\end{document}